# Revised data on γ-families observed in X-ray emulsion chambers of the Experiment PAMIR

V.S. Puchkov, S.E. Pyatovsky[1], R.A. Mukhamedshin, E.A. Kanevskaya, and others

*Lebedev Physical Institute, Russian Academy of Sciences, Leninsky pr. 53, Moscow, 119991 Russia*
*Institute for Nuclear Research, Russian Academy of Sciences, pr 60-letiya Oktyabrya 7a, Moscow, 117312 Russia*

**Abstract.** Recently essential efforts were made to improve measurement routine with X-ray films exposed in the X-ray emulsion chambers at the Pamirs. Analysis of X-ray emulsion response upon recorded events show that γ-family energy and intensity in early publications were over estimated. The main physical results of the new analysis are presented.

## 1 Introduction

Over many years the collaboration of the Experiment PAMIR [1] investigated high-energy nuclear interactions in cosmic rays with X-ray emulsion chambers (XREC) exposed at the Pamirs at an altitude of 4370 m above sea level (600 g/cm$^2$). Various components of nuclear-electromagnetic cascades (NEC) induced by protons and nuclei of primary cosmic rays (PCR) at an energy $E_0 \geq 10^{15}$ eV are recorded. The investigations are concerned mainly with «families», i.e. bundles of the most energetic secondary particles (threshold for particle detection in the XREC is 1–2 TeV) in the core of just the same NEC. The secondaries are hadrons and particles of electromagnetic nature (γ-rays and electrons) which, for brevity, hereafter are referred to as γ-rays.

In this paper we consider families recorded in a thin XREC with lead absorber (Γ-block); its structure schematically is as follows: 4 cm Pb + X-ray film + 1 cm Pb + X-ray film + 1 cm Pb + X–ray film. The XREC of this type were exposed either as a single Γ-block or as an upper part of a compound experimental set-up with hadron blocks. The main fraction of particles recorded in Γ-block are γ-rays, and the corresponding families are called γ-families.

The characteristics of γ-family are sensitive to a composition and energy spectra of the PCR. The analysis is based on comparison of experimental data with simulations.

The proper improvement of measurement procedure with X-ray films exposed in the XREC was made in order to increase an accuracy of physical results.

## 2 Experimental data

The present analysis is based on γ-families recorded during a total XREC exposure ST=2635 m$^2$·year and selected by following criteria: a) total energy of γ-rays in the family $\Sigma E_\gamma \geq 100$ TeV; b) energy of γ-rays and their incidence angle (with respect to vertical) $E_\gamma \geq 4$ TeV and $\theta_\gamma \leq 45°$, respectively; c) deviation of separate γ-rays from the energy-weighted center of a family in the target diagram plane at the observation level $R_\gamma \leq 15$ cm; d) number of γ-rays in the family satisfying above criteria $n_\gamma \geq 3$. The total number of selected and analyzed γ-families is $N_{fam}$=1003.

Among γ-families with total energy $\Sigma E_\gamma \geq 500$ TeV there are events, in which closely related electromagnetic cascades in the central part of γ-family overlap and make up a large diffusive dark spot with high optical density (halo). There are γ-families with single-center and multi-center halo (fig.1).

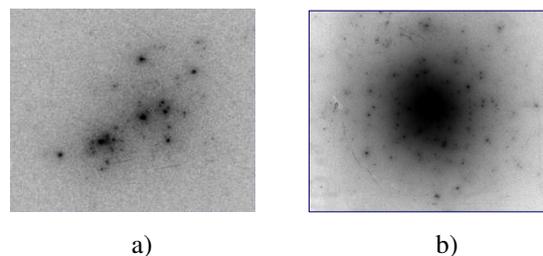

a)          b)

**Fig.1.** a) Example of γ-family on the X-ray film; b) scanner image of the halo event called «FIANIT».

The 61 γ-families with halo were recorded over an exposure ST=3000 m$^2$ year and also used in our analysis. Model calculations show that single-center halo produced

---
[1] Corresponding author.
*E-mail address:* sep@lebedev.ru, vgsep@ya.ru (S.E. Pyatovsky).

mainly by protons, while multicenter halo are produced by heavy nuclei.

The criteria for selection of γ-families with halo are as follows:

a) $\Sigma E_\gamma \geq 500$ TeV;

b) the area of halo $S_{D=0.5}$ bounded by isodense with optical density D=0.5.

$S_{D=0.5} \geq 4$ mm$^2$ – for single-center halo, and $\Sigma S_{i\ D=0.5} \geq 4$ mm$^2$, if $S_{i\ D=0.5} \geq 1$ mm$^2$ – for multi-center halo. Model calculations show that γ-families with $\Sigma E_\gamma = 100$–400 TeV are produced mainly by PCR particles with $E_0 = 10^{15}$–$10^{16}$ eV, while halo γ-families with $\Sigma E_\gamma \geq 500$ TeV are produced by PCR with $E_0 \geq 10^{16}$ eV.

## 3 Model calculations

Artificial events of γ-families were sampled by code MC0 of quark-gluon-string model [2], which was elaborated for Experiment PAMIR and satisfactorily reproduced main characteristics of γ-families with $\Sigma E_\gamma > 100$ TeV. Spectrum of PCR at an energy $E_0 = 2 \cdot 10^{14}$–$3 \cdot 10^{18}$ eV was taken from experimental data of KASKADE and Tibet [3]. Mass composition in MC0 model is presented in Table 1. Calculations revealed that about 80% of all events are produced by primary protons, 10% by helium nuclei and no more than 10% by heavier nuclei. This conclusion is almost independent of the models and thus provides a possibility to estimate the fraction of protons in PCR in the range of $E_0 = 10^{15}$–$10^{17}$ eV.

**Table 1.** Mass composition in MC0 model.

| E, eV | $10^{15}$ | $10^{16}$ | $10^{17}$ |
|---|---|---|---|
| p, % | 33 | 26 | 20 |
| α, % | 22 | 17 | 15 |

Detailed analysis of a response of the XREC to events recorded X-ray films is based on simulation of measurement procedure by code GEANT3.21. The contribution of under threshold electromagnetic cascades, the mutual influence of neighboring cascades in γ-family, identification of cascades against background by Raleigh criterion are taken into account. The analysis show that in our early publication there was overestimation of family energy approximately by 20%.

## 4 Spatial distribution and structure of γ-families

Analysis of spatial distribution in γ-families show that due to fluctuation, a large number of γ-families with $\Sigma E_\gamma$ close to threshold ($\Sigma E_{\gamma thr} = 100$ TeV) have broader spatial distributions than average one (with $R_f = \Sigma R_\gamma / n_\gamma$ larger by 15%). Experimental value of $<R_f>$ for range of $\Sigma E_\gamma = 100$–400 TeV is $<R_f> = 1.94 \pm 0.06$ cm, while model values $<R_f^{MC0}> = 2.01 \pm 0.03$ cm, $<R_f^P> = 1.79 \pm 0.03$, $<R_f^{He}> = 2.37 \pm 0.09$, $<R_f^{Fe}> = 4.15 \pm 0.18$ cm.

Comparison $<R_f>$ and $<R_f^{MC0}>$ indicates that experimental γ-families at energy range $E_0 = 10^{15}$–$10^{16}$ are produced mainly by protons. At the same time contribution of helium nuclei less than 15%.

To prove the suggestion that at energy $E_0 > 10^{16}$ eV most events with halo are generated by primary protons we compare the fraction of γ-families with multi-center halo in the experiment and calculations (Table 2).

**Table 2.** Fraction of γ-families with multi-center halo generated by different nuclei.

| p | He | C | Fe | PAMIR |
|---|---|---|---|---|
| 0.25 ±0.03 | 0.45 ±0.09 | 0.59 ±0.11 | 0.70 ±0.12 | 0.23 ±0.07 |

It is evident that recorded γ-families with halo almost entirely are generated by primary protons with possible little addition of He.

## 5. Dependence of g-family flux on the PCR mass composition

The intensities of the γ-families in MC0 model and Pamir experiment are presented in Table 3.

**Table 3.** Energy range: $10^{15} \div 10^{16}$ eV.

| | MC0 | PAMIR* (with chamber response) |
|---|---|---|
| $I_f(\Sigma E_g > 100$ TeV, 600 g/cм$^2$, θ=0), m$^{-2}$·year$^{-1}$·sr$^{-1}$ | 0.71 ± 0.02 | 0.39 ± 0.03 |

To match this experimental value of γ-families intensity in the interval of $E_0 = 10^{15}$-$10^{16}$ eV and calculations of code MC0 fraction of protons in the model should be reduced from 33% to (18±2)% if the ejected part of protons is substituted by nuclei of iron group or to (16±2)% if the ejected protons are substituted by helium nuclei. The similar conclusion with the same quantitive estimation of the proton fraction in PCR was made for the energy $E_0 > 10^{16}$ eV by analysis of γ-families with halo.

Thus proton fraction in PCR in the energy range $10^{15} \div 10^{17}$ eV in Pamir experiment is about (16 ÷ 18)%, Table 4.

**Table 4.** Fraction of protons in the energy range $10^{15} \div 10^{17}$ eV.

| MC0 | PAMIR* (allowing for chamber response) |
|---|---|
| 33% | (16 ÷ 18)% ± 2% |

## 6. Conclusions

1) γ-families in the Experiment PAMIR at an primary energy $E_0 = 10^{15}$-$10^{16}$ eV are produced by primary protons with a small amount (~10%) of helium;

2) the fraction of protons in the PCR composition at an energy $E_0 = 10^{15}$-$10^{16}$ eV is about 15-20% and do not change appreciably up to $E_0 = 10^{17}$ eV.